\documentclass[twocolumn,aps,prl,amssymb,amstext,showpacs]{revtex4}

\usepackage{graphicx,amsfonts,amssymb,amsmath}
\usepackage{epsfig}
\usepackage{color}

\newcommand{\ip}[2]{\left\langle #1, #2 \right \rangle}
\newcommand{\half}{\mbox{$\frac{1}{2}$}}
\newcommand{\xhat}{\hat{{\bf x}}}
\newcommand{\yhat}{\hat{{\bf y}}}
\newcommand{\e}{{{\epsilon}}}
\newcommand{\beq}[0]{\begin{equation}}
\newcommand{\eeq}[0]{\end{equation}}
\newcommand{\R}{\mathbb{R}}
\newcommand{\mvec}{{\mathbf m}}
\newcommand{\Mvec}{{\mathbf M}}
\newcommand{\brackets}[1]{\left( #1 \right)}
\newcommand{\sgn}{\operatorname{sgn}}
\newcommand{\avg}[1]{\left< #1 \right>} 
\newcommand{\mvecp}{{\mathbf m_+}}
\newcommand{\mvecm}{{\mathbf m_-}}

\newcommand{\pdd}[2]{\frac{\partial^2 #1}{\partial #2^2}}

\DeclareMathOperator{\sech}{sech}

\begin{document}

\title{Fast domain wall propagation in uniaxial nanowires with transverse fields}


\author{Arseni Goussev$^{1,2}$, Ross G.~Lund$^3$,  JM Robbins$^3$, Valeriy Slastikov$^3$, Charles Sonnenberg$^3$}

\affiliation{
$^1$Department of Mathematics and Information Sciences, Northumbria University, Newcastle Upon Tyne, NE1 8ST, UK\\
$^2$Max Planck Institute for the Physics of Complex
  Systems, N{\"o}thnitzer Stra{\ss}e 38, D-01187 Dresden,
  Germany\\ 
 $^3$School of Mathematics, University of Bristol,
  University Walk, Bristol BS8 1TW, United Kingdom}

\date{\today}

\begin{abstract}
Under a magnetic field along its axis, domain wall motion in a  uniaxial   nanowire is much slower than in the fully anisotropic case,  typically by several orders of magnitude (the square of the dimensionless Gilbert damping parameter).  
However, with the  addition of a magnetic field transverse to the wire, this behaviour is dramatically reversed;  up to a critical field strength, analogous to the Walker breakdown field,  domain walls in a uniaxial wire 
propagate faster than in a fully anisotropic wire { (without transverse field)}.   Beyond this critical field strength, precessional motion sets in, and the mean velocity decreases.  Our results are based on leading-order analytic calculations of the velocity and critical field as well as numerical solutions of the Landau-Lifshitz-Gilbert equation.  \end{abstract}

\pacs{75.75.-c, 75.78.Fg}

\maketitle




\subsection{Introduction} The dynamics of magnetic domain walls  in ferromagnetic nanowires under external magnetic fields  \cite{SchryerWalker74, Yamaguchi04, Allwood05, Cowburn07, Beach05, Yang08, Bryan08,  Tretiakov08, Wang09, Wang_etal09, Lu10, Mougin07} 
and spin-polarised currents \cite{Mougin07, Li04, Thiaville05, Beach06,  Parkin08, Hayashi08, Thomas10, Tretiakov10, Tretiakov12}  is  a central problem in micromagnetics and spintronics, both as a basic physical phenomenon as well as  a cornerstone of magnetic memory and logic technology \cite{Allwood05, Parkin08, Hayashi08, Thomas10}.  From the point of view of applications, it is desirable to maximise the domain wall velocity in order to optimise switching and response times. 

 Partly because of fabrication techniques, attention has been focused on nanowires with large cross-sectional aspect ratio, typically of rectangular cross-section.  In this case, even if the bulk material is isotropic (e.g., permalloy), the domain  geometry induces a fully anisotropic magnetic permeability tensor, with easy axis along the wire and hard axis along its shortest dimension  \cite{SS, HubertSchaefer98}.  Nanowires with uniaxial permeability, characteristic of more symmetrical cross-sectional geometries  (e.g., square or circular), have been less studied \cite{Sun10, our_PRL, Y_Gou_paper}.  {\color{black} Here we investigate domain wall (DW) motion in uniaxial wires in the presence of transverse fields. We show that the DW velocity in uniaxial wires depends strongly on the longitudinal applied field $H_1$, increasing with $H_1$ up to a certain critical field and thereafter falling off as precessional motion sets in.}  {\color{black} We employ a systematic asymptotic expansion scheme, which differs from alternative approaches based on approximate dynamics for the DW centre and orientation; a detailed account of this scheme, also including anisotropy and current-induced torques, will be given separately \cite{paper1} . }

We  employ a continuum description of the magnetisation.  For a thin nanowire, this  is provided by the one-dimensional Landau-Lifshitz-Gilbert (LLG) equation \cite{LandoLifshitz35, Gilbert55, Kosevich90, HubertSchaefer98}, which we write in the non-dimensionalised form
{\color{black}
 \begin{equation}\label{eq: LL}
{\mathbf{\dot M}} = \gamma \mathbf{M}\times \mathbf{H } - \alpha
\mathbf{M}\times \left( \mathbf{M}\times \mathbf{H } \right).
\end{equation}
 Here $\mathbf{M}(x,t)$  is a unit-vector field specifying  the orientation of the magnetisation, which we shall also 
 write in polar form ${\mathbf{ M}} = (\cos\Theta,\sin\Theta\cos\Phi,\sin\Theta\sin \Phi)$. The effective magnetic field, $\mathbf{H }(\mathbf{m} )$, is
given by
\begin{equation}
\label{ eq: H(m)}
\mathbf{H } = A\mathbf{m'' } + K_1 m_1 \xhat - K_2 m_2 \yhat  + \mathbf{H_a }.
\end{equation} 
Here  $A$ is the exchange constant, $K_1$ is the easy-axis anisotropy, $K_2 > 0$  is the hard-axis anisotropy, $\mathbf{H_a }$  is the applied magnetic field (taken to be constant), $\gamma$ is the gyromagnetic ration, and $\alpha$ is the Gilbert damping
parameter.  For convenience we choose units for length, time and energy so that $A = K_1 = \gamma = 1$.} 
Domains correspond to locally uniform configurations in which $\Mvec$ is aligned along one of the local minima,  denoted $\mvecp$ and $\mvecm$,   of the potential energy
\begin{equation}
\label{eq: U }
 U (\mvec) = -\half  (m_1^2 - K_2 m_2^2) -   \mathbf{m}\cdot \mathbf{H_a}.
\end{equation}
Two distinct domains separated by a DW are described by the boundary conditions  $\mathbf{M}(\pm \infty,t) = \mvec_{\pm}$.

For purely longitudinal fields $ \mathbf{H_a} = H_1 \xhat$ and  for $H_1$ below  the Walker breakdown field  $H_W = \alpha K_2/2$, the DW propagates as a travelling wave \cite{SchryerWalker74},
the so-called Walker solution $\Theta(x,t) = \theta_W(x-V_Wt)$,  $\Phi(x,t) = \phi_W$, where $\theta_W$ and $\phi_W$ are given by
\begin{equation}
\label{eq: theta_W }
 \theta_W(\xi) = 2\tan^{-1} 
 ( e^{-\xi/\gamma}
 ), \ \ \sin 2\phi_W = H_1/H_W.
 \end{equation}
 The width of the DW, $\gamma$,  is given by  $\gamma = ( 1 + K_2\cos^2\phi_W )^{-1/2}$, and the velocity is given by
\begin{equation}
\label{ eq: V_W}
V_W = -\gamma (\alpha+1/\alpha) H_1.
\end{equation}
 For $H_1 > H_W$,  the DW undergoes non-uniform  precession and translation, with mean velocity decreasing with $H_1$ \cite{SchryerWalker74, Beach05, Yang08, Wang09}.  The  effects of additional transverse fields  have been examined  recently \cite{Bryan08, Lu10}.

If the cross-sectional geometry is sufficiently symmetrical (e.g., square or circular), the permeability tensor becomes uniaxial,  so that  $K_2 = 0$ \cite{SS, HubertSchaefer98}. 
The dynamics in this case  is strikingly different.  The LLG equation has an exact  solution, $\Theta(x,t) = \theta_0(x - V_P t)$,  $\Phi(x,t)  = -H_1 t$, in which the DW propagates with velocity
\begin{equation}
\label{ eq: V_P}
V_P = -\alpha H_1
\end{equation}
and precesses about the easy axis with angular velocity $-H_1$ \cite{Sun10, our_PRL}.  The precessing solution persists for all $H_1$ -- there is no breakdown field -- but becomes unstable for $H_1 \gtrsim 1/2$ \cite{Y_Gou_paper}.  

For $H_1 < H_W$, the ratio $V_W/V_P$ is given by $\gamma(\alpha^{-2} + 1)$.  For typical values of $\alpha$ ($ 0.01$ -- $0.1$), the uniaxial velocity $V_P$ is less than the fully anisotropic velocity $V_W$ by several orders of magnitude. 
As we show below,  applying a transverse field $H_2 > 0$ to a uniaxial wire dramatically changes its response to an applied longitudinal field $H_1$.  The transverse field, analogous to  hard-axis anisotropy, inhibits precession and facilitates fast DW propagation.
 {\color{black} For $H_1$ less than an $H_2$-dependent critical field $H_{1c}$, given in the linear regime by \eqref{eq: h_1c} below, there appears a travelling wave, while for $H_1 >H_{1c}$, there appears an oscillating solution, as in the Walker case.  The DW velocity of travelling wave exceeds that of oscillating solution.  }

\subsection{Velocity of travelling wave}
We first obtain a  general identity, of independent interest, which relates the velocity of a travelling wave  
$\mathbf{M } (x,t) = \mathbf{m} (x-Vt)$ (assuming one exists) 
to the change in potential energy across the profile (for zero transverse field, this coincides with results of  \cite{SchryerWalker74} and \cite{Wang_etal09}).
Noting that $\mathbf{\dot M} = -V \mathbf{ m'}$, we take the square of \eqref{eq: LL} and integrate over the length of the wire to obtain
\begin{equation}
\label{ eq: velocity identity 1}
V^2 
||{\mathbf{m'}} ||^2
 = (1 + \alpha^2) 
||\mathbf{m}\times \mathbf{H}||^2.
\end{equation}
Here we use the notation
\begin{equation}
\label{eq: norm and inner product }
 || \mathbf{u}||^2 = \ip{\mathbf{u}}{\mathbf{u}}, \quad \ip{\mathbf{u}}{\mathbf{v}} = \int_{-\infty}^\infty {\mathbf{u}}\cdot {\mathbf{v}}\, dx
\end{equation}
for the $L^2$-norm and inner product of vector fields (analogous  notation for  scalar fields is used below).
Next, we take the inner product of  \eqref{eq: LL} with $ \mathbf{H}$ to obtain
\begin{equation}
\label{ eq: velocity identity 2}
V \ip{ \mathbf{ m'}} {\mathbf{H} }  = - \alpha ||\mathbf{m}\times \mathbf{H}||^2.
\end{equation}
Noting that $\mathbf{ m'}\cdot \mathbf{H} = \left(\half  \mathbf{m'}\cdot \mathbf{m'} - U(\mathbf{m})\right)' $, 
we combine \eqref{ eq: velocity identity 1} and \eqref{ eq: velocity identity 2} to obtain
\begin{equation}
\label{eq: velocity profile final }
V = \half (\alpha + 1/\alpha)
\, ||\mathbf{m'}||^{-2}\,  \left(U(\mathbf{m_-}) - 
U(\mathbf{m_+} )\right).
\end{equation}
The identity \eqref{eq: velocity profile final } has a simple physical interpretation; 
the velocity is proportional to the potential energy difference 
across the wire, and 
inversely proportional to the exchange energy of the profile.

From now on, we consider the uniaxial case $K_2 = 0$  and applied field with longitudinal and transverse components $H_1, H_2 > 0$ (by symmetry, we can assume $H_3 = 0$) with $|\mathbf{H_a}| < 1$.  An immediate consequence of \eqref{eq: velocity profile final } is that, in the uniaxial case,  the  velocity must vanish as $H_1$  goes to zero.  For when $H_1 = 0$,  the local minima ${\mathbf m}_\pm$ are  related by reflection through the $23$-plane, and $U({\mathbf m}_+) = U({\mathbf m}_-)$.

%
%

\subsection{Small transverse field}
{\color{black}

In order to understand travelling wave and oscillating solutions as well as the transition between them, we first carry out an asymptotic analysis in which both $H_1$ and $H_2$ are regarded as small, writing $H_1=\e h_1$, $H_2=\e h_2$ and rescaling time as $\tau =\e t$ (a systematic treatment including current-induced torques will be given in \cite{paper1}). We seek a solution of  the LLG equation  \eqref{eq: LL} of the following asymptotic form:
\begin{align}
\Theta (x,t) =\theta_0 (x,\tau) + \e \theta_1 (x,\tau) + ..., \label{eq: Theta expand 1}\\
\Phi(x,t) = \phi_0 (x,\tau) + \e \phi_1 (x,\tau) + ...\label{eq: Phi expand 1}
\end{align}
It is straightforward to check that the boundary conditions, namely that $\mathbf{m}$ approach distinct minima of $U$ as $x\rightarrow \pm \infty$, imply that
\begin{equation} \label{BC1}
\mathbf{m} (\pm \infty, \tau) = (\pm 1, \e h_2, 0) + O(\e^2).
\end{equation}

The leading-order equations for $\Theta$ and $\Phi$ become
\begin{align}
\theta_{0,xx}-\half(1+\phi_{0,x}^2)\sin 2 \theta_0 &= 0,\label{LOS1} \\
\brackets{\sin^2\theta_0 \phi_{0,x}}_x&=0. \label{LOS2}
\end{align}
The only physical (finite-energy) solutions of \eqref{LOS1} and \eqref{LOS2} consistent with the boundary conditions \eqref{BC1} are of the form 
\begin{gather}
\phi_0(x,\tau)=\phi_0(\tau)\label{eq: phi_0 first}\\
\theta_0(x,\tau)=2\arctan \exp(-(x-x_*(\tau))),\label{eq: theta_0 first}
\end{gather}
 where $\phi_0$ and $x_*$ respectively describe the DW orientation and centre, and are  functions of $\tau$ alone.   
It is convenient to introduce a travelling coordinate $\xi=x-x_*(\tau)$ and rewrite the ansatz \eqref{eq: Theta expand 1}--\eqref{eq: Phi expand 1}  as
\begin{align}
\Theta (x,t) =\theta_0 (\xi,\tau) + \e \theta_1 (\xi,\tau) + ...\, , \\
\Phi(x,t) = \phi_0 (\xi,\tau) + \e \phi_1 (\xi,\tau) + ...
\end{align}

To obtain equations for $\phi_0(\tau)$ and $x_*(\tau)$ we must proceed to the next order.
%
It is convenient to introduce new  variables  at order $\epsilon$ which, in light of the boundary conditions \eqref{BC1}, vanish at $  x=\pm \infty$, as follows: 
\begin{align}
\Theta_1:= \theta_1 -  h_2 \cos \phi_0 \cos \theta_0 ,\\
u:=\phi_1 \sin\theta_0 +   h_2 \sin \phi_0.
\end{align}
These satisfy 
the   linear inhomogeneous equations
\begin{align}
L\Theta_1 &= f,  
\label{eq: LinSysTheta}  \\
Lu &= g.
\label{eq: LinSysu}
\end{align}
Here $L$ is the self-adjoint Schr\"odinger operator given by
\beq \label{eq: L}
L=-\pdd{}{\xi}+ W(\xi), 
\eeq
where
\beq W = \frac{\theta_0'''}{\theta_0'} = 1 - 2\sech^2\xi,  \label{eq: W}
\eeq
and $f(\xi,\tau)$ and $g(\xi,\tau)$ are given by
\begin{gather}
f
={(1+\alpha^2)}^{-1}\sin\theta_0(-\alpha \dot{x}_* -\dot{\phi}_0 )- h_1\sin\theta_0, \nonumber\\
g
={(1+\alpha^2)}^{-1}\sin\theta_0
(\dot{x}_* -\alpha\dot{\phi}_0 )
 + 2 h_2 
\sin^2\theta_0 \sin\phi_0.
\end{gather}

The DW position $x_*$ and orientation $\phi_0$ are determined from the solvability conditions for \eqref{eq: LinSysTheta} -- \eqref{eq: LinSysu}.
According to the Fredholm alternative,  given a self-adjoint operator $L$ on $L^2(\R)$, a necessary condition for the equation $L \Theta_1= f$ to have a solution $\Theta_1$ is that $f$ be orthogonal to the kernel of $L$. If this is the case, a sufficient condition is that the spectrum of $L$ is isolated away from $0$.   From \eqref{eq: L} and \eqref{eq: W} it is clear that $\theta_0'$ belongs to the kernel of $L$, and since the eigenvalues of a one-dimensional Schr\"odinger operator are nondegenerate, it follows that $\theta_0'$ spans the kernel of $L$.  Moreover, since  $W(\xi) \rightarrow 1$  as $\xi \rightarrow \pm \infty$, it follows that the spectrum of $L$ is discrete near $0$.  (In fact, $W$ is a special case of the exactly solvable P\"oschl-Teller potential, but we won't make use of this fact.)
Requiring 
$f$ and $g$ in  \eqref{eq: LinSysTheta} and \eqref{eq: LinSysu}  to be orthogonal to $\theta_0'$ and noting that $\ip{\theta_0'}{\theta_0'} = 2$, $\ip{\theta_0'}{\sin \theta_0} = -2$, $\ip{\theta_0'}{1} = -\pi$, and 
$\ip{\theta_0'}{\cos \theta_0} = 0$, we  obtain the following system of ODEs for $\phi_0$ and $x_*$:
\begin{align}
\dot{\phi}_0 &= -h_1 - \frac{\alpha \pi}{2} h_2 \sin \phi_0,
\label{SolConPhi}\\
\dot{x}_*&=-\alpha h_1 +  \frac{\pi}{2} h_2 \sin \phi_0. \label{SolConx}
\end{align}

Travelling wave solutions appear provided \eqref{SolConPhi} has fixed points; this occurs for $h_1$ below a critical field 
 $h_{1,c}$ given by
\begin{equation} \label{eq: h_1c}
h_{1,c} =\frac{\alpha \pi h_2}{2}, 
\end{equation}
The velocity and orientation of the travelling wave are given by
\begin{gather}
 \dot{x}_* = -\left(\alpha+\frac{1}{\alpha}\right) h_1, \label{eq: x_*dot}\\
\sin \phi_0  = -\frac{h_1}{h_{1,c}}.  \label{eq: phi_0dot}
\end{gather}
There are two possible solutions for $\phi_0\in [0,2\pi)$, only one of which is stable. 
Oscillating solutions appear for $h_1 > h_{1c}$, and are given by
\begin{equation}
h_1 \tan\half \phi_0 
= -h_{1,c} 
- \sqrt{h_1^2 - h_{1,c}^2}
\tan \brackets{\half \sqrt{h_1^2 - h_{1,c}^2}
\, \tau}
\end{equation}
with the period  $T=2\pi/\sqrt{h_1^2 - h_{1,c}^2}$. The mean precessional and translational velocities are obtained  by averaging over a period, with result
\begin{align}
\avg{\dot{\phi}_0}&=-\sgn(h_1)\sqrt{h_1^2-h_{1,c}^2},\label{eq: mean precession}\\  
\avg{\dot{x}_*}&= -\left(\alpha+\frac{1}{\alpha}\right)h_1 + \frac{1}{\alpha}\sgn(h_1) \sqrt{h_1^2-h_{1,c}^2}.\label{eq: mean velocity}
\end{align}
Note that for $h_1 = h_{1,c}$,  \eqref{eq: mean velocity} coincides with the travelling wave velocity \eqref{eq: x_*dot}, whereas for $h_1 \gg h_{1,c}$,  \eqref{eq: mean velocity} reduces to the velocity of the precessing solution given by \eqref{ eq: V_P}.

The behaviour is similar in many respects to the Walker case (i.e., $K_2 \neq 0$ and $H_2 = 0$).  Here, the transverse field rather than hard-axis anisotropy serves to arrest the precession of the DW (provided the longitudinal field is not too strong).  There are differences as well; in the transverse-field case there is just one stable travelling wave, whereas in the Walker case there are two.  Also, in the transverse-field case the asymptotic value of the magnetisation has a transverse component, whereas in the Walker case it has none.}

\subsection{Moderate transverse field}
We can  extend  the travelling wave analysis to the regime where $H_2$ is no longer regarded as small.  We continue to regard $H_1$ as small, writing $H_1 = \epsilon h_1$ and $V = \epsilon v$, and expand the travelling wave ansatz $\Theta(x,t) = \theta(x-Vt)$, $\Phi(x,t) = \phi(x-Vt)$  to first order in $\epsilon$, writing $\theta = \theta_0 + \epsilon \theta_1$, $\phi = \phi_0 + \epsilon \phi_1 $.  Substituting into the LLG equation, we obtain  the $O(\epsilon^0)$ equations 
\begin{equation}
\label{eq: theta_0 H_2 }
\theta_0' = (H_2 - \sin\theta_0), \quad \phi_0 = 0,
\end{equation}
with boundary conditions $  \sin \theta_{0\pm} = H_2$, $\theta_{0+} > \pi/2$ and $\theta_{0-} <   \pi/2$.  
Thus, for $H_2 = O(\epsilon^0)$,  azimuthal symmetry is broken at leading order, and the static  profile is parallel to the transverse field  (the alternative solution with  $\phi_0 =\pi$ is unstable).  The solution of \eqref{eq: theta_0 H_2 } is given  by 
 \begin{multline}
\tan \frac{ \theta_0}{2} = 
\frac{\kappa}{H_2} \tanh \left[
    \tanh^{-1} \left( \frac{H_2 - 1}{\kappa} \right) -
    \frac{\kappa}{2}\xi \right] + \frac{1}{H_2}
\label{eq:theta0},
\end{multline}
where $\kappa = \sqrt{1-H_2^2}$.

At order $\epsilon$ we obtain the linear inhomogeneous equations
\begin{align}
   L \theta_1 &=    \frac{\alpha}{1+\alpha^2}  v  \theta_0' - h_1 \sin\theta_0, \label{eq: linear theta}\\
    M  \phi_1 &=   \frac{1}{1+\alpha^2}  v (\cos \theta_0)',  \label{eq: linear phi}
\end{align}
where 
\begin{equation}
\label{eq: linear ops }
L
= -\frac{d^2}{d\xi^2}  + \frac{\theta_0'''}{\theta_0'}, \ \ M
 = -\frac{d}{d\xi} \sin^2\theta_0 \frac{d}{d\xi} + H_2\sin\theta_0.
\end{equation}
Here $\theta_0$ is given by \eqref{eq:theta0},
and  $\theta_1$, $\phi_1$ are required to vanish as $\xi\rightarrow \pm \infty$.  
As above, the Fredholm alternative implies that the right-hand side of  \eqref{eq: linear theta} must be orthogonal to $\theta_0'$ in order for a solution to exist. Calculation yields
\begin{equation}
\label{eq: linear v }
V = -\left(\alpha+\frac{1}{\alpha}\right) (1 - (H_2 / \kappa) \cos^{-1}H_2)^{-1} H_1.
\end{equation}
For $H_2 = 0$, this coincides with \eqref{eq: x_*dot}; thus, \eqref{eq: linear v }  gives  $H_2$-nonlinear corrections to the velocity.  Moreover,  it is straightforward to show that \eqref{eq: linear v }  is consistent with the general identity \eqref{eq: velocity profile final }.  Finally, one can also show that $M$ has trivial kernel with spectrum bounded away from zero, so that \eqref{eq: linear phi}  is automatically solvable.

It is interesting to compare the DW velocity with  transverse field to the Walker case.  From \eqref{ eq: V_W} and \eqref{eq: linear v },\begin{equation}
\label{eq: uni/Walker velocity ratio }
V_W/V= \gamma\, (1 - (H_2 / \kappa) \cos^{-1}H_2) < 1.
\end{equation}
Thus, to leading order in $H_1$,  the DW velocity in a uniaxial wire with transverse field exceeds the Walker velocity.  Numerical results below establish that this continues to hold 
as $H_1$ approaches the critical field $H_{1c}$.

\subsection{Numerical results}
To verify our analytical results, we solve the LLG equation \eqref{eq: LL} using a finite-difference scheme on
a  domain $-L \le x \le L$ where $L =100$ (the DW has width of order 1).  Neumann boundary conditions, $\mathbf{m'} = 0$, are maintained at the endpoints.  The  damping parameter $\alpha$ is taken to be $0.1$ throughout. As  initial condition we take the stationary 
profile, with $\theta_0$ given by \eqref{eq:theta0} and $\phi_0 = 0$.   After an initial transient period, during which the asymptotic values  of $\mathbf{m}$ at $x\rightarrow \pm L$ converge to $\mathbf{m_\pm}$,
a stable solution emerges, in which the DW propagates with a characteristic mean velocity $V$.  (For convenience, we have taken $H_1 < 0$, so that $V$ is positive.)  {\color{black} In Figure~\ref{fig1}, numerically computed values of $V$ are plotted as a function of $|H_1|$
 for three fixed values of the transverse field: $H_2
= 0.2$, $H_2 = 0.1$, and the limiting case $H_2 = 0$, where the
dynamics  is given by the precessing solution.  There is good quantitative agreement with the analytic results for small transverse fields, \eqref{eq: x_*dot}, for $|H_1| < H_{1,c}$, and \eqref{eq: mean velocity},  for $|H_1| > H_{1,c}$, 
In Figure~\ref{fig2}, the analytic expressions for the velocity for small and moderate transverse fields are compared to numerical results for $H_2 = 0.2$ and $|H_1| \ll H_{1c}$.  The moderate-field expression \eqref{eq: linear v }, which depends nonlinearly in $H_2$, gives excellent agreement for small driving fields.
For nonzero $H_2$, 
the velocity 
exhibits a peak at a critical field 
$|H_{1\mathrm{c}}|$, which depends on $H_2$.
\begin{figure}[ht]
\includegraphics[width=3.4in]{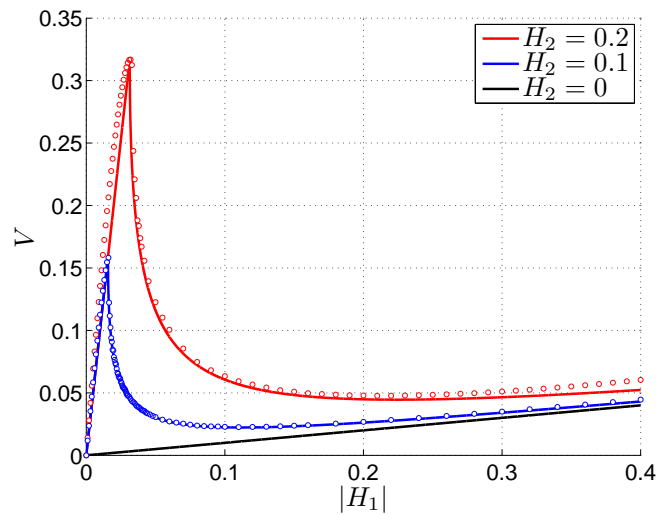}
\caption{\color{black} Average DW velocity $V$ as a
  function of the driving field  $|H_1|$ for three 
  values of the transverse field $H_2$.  The analytic formulas (solid curves) \eqref{eq: x_*dot}, for $|H_1| < H_{1,c}$, and \eqref{eq: mean velocity},  for $|H_1| > H_{1,c}$, are plotted against numerically computed values (open circles).  For $H_2 = 0$, the analytic formula is exact.
 }\label{fig1}
\end{figure}
\begin{figure}[ht]
\includegraphics[width=3.4in]{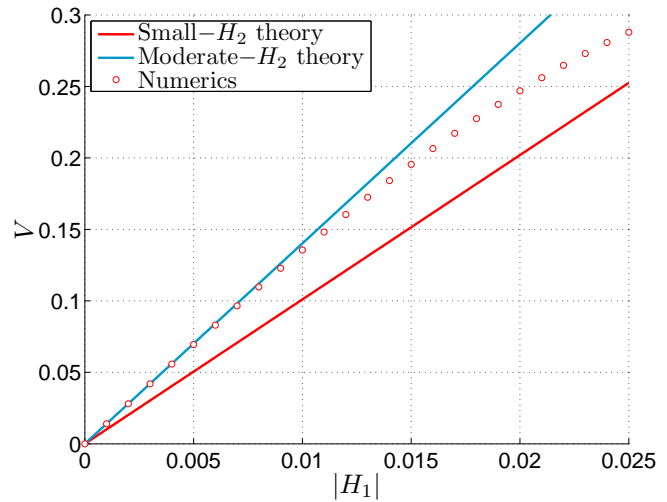}
\caption{\color{black} DW velocity $V$ as a
  function of the driving field  $|H_1|$ for  $H_2 = 0.2$.  The expressions for  small-transverse field  \eqref{eq: x_*dot}  (red curve) and moderate-transverse field  \eqref{eq: linear v }
(light blue curve) 
 are plotted against numerically computed values (open circles).  
 }\label{fig2}
\end{figure}
Figure~\ref{fig3} shows the dependence of the critical field $|H_{1,\mathrm{c}}|$
on $H_2$, in close agreement with the analytic result \eqref{eq: h_1c}. 
\begin{figure}[ht]
\includegraphics[width=3.4in]{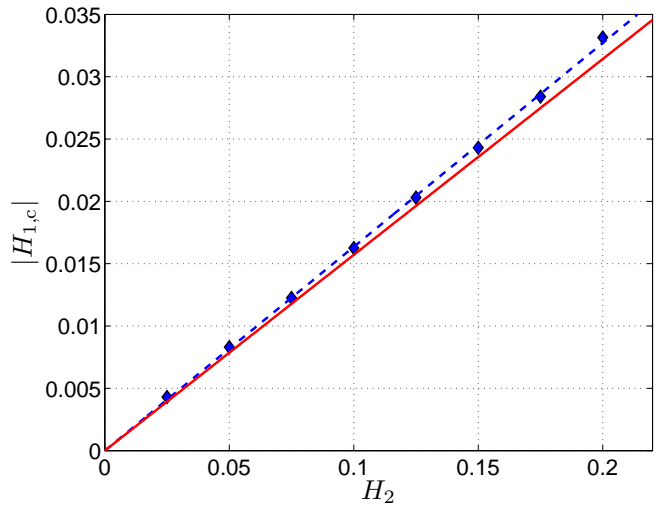}
\caption{\color{black} The critical driving field
  $|H_{1,\mathrm{c}}|$ as a function of the transverse field
  $H_2$. A linear fit (blue curve) through the numerically computed data (blue diamonds) is 
  plotted alongside the  
 analytical result \eqref{eq: h_1c} (red curve).}
\label{fig3}
\end{figure}
\begin{figure}[ht]
\includegraphics[width=3.5in]{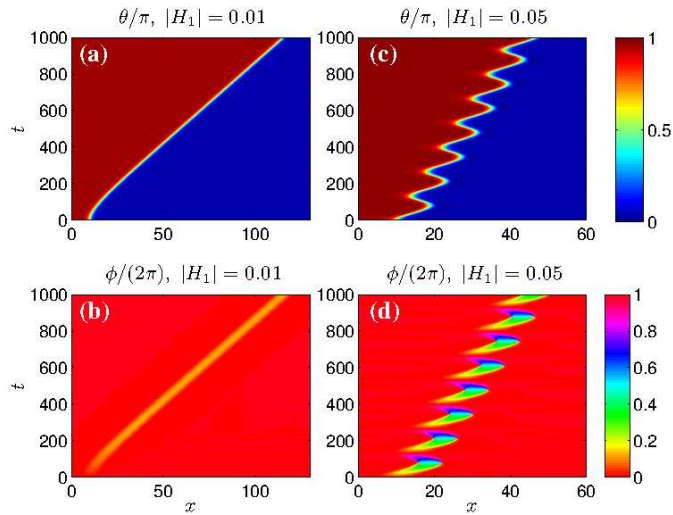}
\caption{The magnetization distribution, 
  $\theta(x,t)$ and $\phi(x,t)$, for two values of the driving field:
  $H_1 = -0.01$ in figures (a) and (b), and $H_1 = -0.05$ in figures (c)
  and (d). The transverse field is taken as $H_2 = 0.1$ throughout.}
\label{fig4}
\end{figure}}

As in the Walker case, the properties of the propagating solution are qualitatively different for driving fields $|H_1|$ below and above the critical field. This is confirmed in
Figure~\ref{fig4}, which shows contour plots of the  magnetization in the $(x,t)$-plane. 
Figs.~\ref{fig4}(a) and
\ref{fig4}(b), where $H_1 = -0.01$,  exemplify the case  $|H_1|  < |H_{1\mathrm{c}}|$.
The magnetisation evolves as a fixed profile translating rigidly with velocity $V$. 
For $|H_1| > |H_{1\mathrm{c}}|$, as exemplified by
Figs.~\ref{fig4}(c) and \ref{fig4}(d),  where $H_1 =
-0.05$. 
the magnetization
profile exhibits a  non-uniform precession as it propagates
along the nanowire,  with mean velocity in good agreement with \eqref{eq: mean velocity}.


\subsection{Summary}
We have established, both analytically in leading-order asymptotics and numerically, the existence of travelling wave and oscillating solutions of the LLG equation in uniaxial wires in applied fields with longitudinal and transverse components.  We have obtained analytic expressions for the velocity, \eqref{eq: x_*dot} and \eqref{eq: linear v }, and for the critical longitudinal field, \eqref{eq: h_1c}, above which the travelling wave solution ceases to exist.  We have also obtained the mean precessional and linear velocities \eqref{eq: mean precession} and \eqref{eq: mean velocity} for oscillating solutions. The analytic results are confirmed by  numerics.  
%
%

%


\begin{thebibliography}{99}

 \bibitem{SchryerWalker74} N.~L.~Schryer and L.~R.~Walker,
  J. Appl. Phys. {\bf 45}, 5406 (1974).
  
 \bibitem{Yamaguchi04} A.~Yamaguchi, T.~Ono, S.~Nasu, K.~Miyake, K.~Mibu, T.~Shinjo, Phys.~Rev.~Lett.~{\bf 92} 077205 (2004).

\bibitem{Allwood05} D.A.~Allwood, G.~Xiong, C.C.~Faulkner,
  D.~Atkinson, D.~Petit and R.P.~Cowburn, Science {\bf 309}, 1688
  (2005).
  

\bibitem{Cowburn07} R.P.~Cowburn, Nature (London) {\bf 448}, 544
  (2007).
  
  \bibitem{Beach05} G.S.D~Beach, C.~Nistor,  C.~Knutson, M.~Tsoi, and J.L.~Erskine, Nature~Mater.~{\bf 4}, 741(2005).



  
  \bibitem{Yang08}J.~Yang, C.~Nistor, G.S.D.~Beach, and J.L.~Erskine, Phys.~Rev.~B {\bf 77}, 014413 (2008).
  
  
  \bibitem{Bryan08} M.T. ~Bryan, T.~Schrefl, D.~Atkinson, D.A.~Allwood, J.~Appl.~Phys.~{\bf 103}, 073906 (2008).
  
  
  \bibitem{Tretiakov08} O.A.~Tretiakov, D.~Clarke, Gia-Wei Chern, Ya.~B.~Bazaliy and O.~Tchernyshyov, Phys.~Rev.~Lett.~{\bf 100} 127204 (2008).


\bibitem{Wang09}X.R.~Wang, P.~Yan, J.~Lu, Europhys.~Lett.~{\bf 86}, 67001
(2009).  



    \bibitem{Wang_etal09} X.R.~Wang, P.~Yan , J.~Lu, C.~He, Ann.~Phys. {\bf 324}, 1815--1820 (2009).




  
  \bibitem{Lu10} J.~Lu and X.R.~Wang, J.~Appl.~Phys.~{\bf 107}, 083915 (2010).
  


\bibitem{Mougin07}A.~Mougin, M.~Cormier, J.P.~Adam, P.J.~Metaxas and J.~Ferr\'e, Europhys.~Lett.~{\bf 78}, 57007 (2007).
  
  \bibitem{Li04} Z.~Li and S.~Zhang, Phys.~Rev.~Lett.~{\bf 92} 207203 (2004).
  
  
  \bibitem{Thiaville05}A.~Thiaville, Y.~Nakatani,, J.~Miltat and Y.~Suzuki, Europhys.~Lett.~{\bf 69}, 990 (2005).


\bibitem{Beach06}G.S.D~Beach, C.~Knutson, C.~Nistor, M.~Tsoi, and J.L.~Erskine, Phys.~Rev.~Lett.~{\bf 97} 057203 (2006)


  
\bibitem{Parkin08} S.~S.~P.~Parkin, M.~Hayashi and L.~Thomas, Science
  {\bf 320}, 190 (2008).
  
  \bibitem{Hayashi08} M.~Hayashi, L.~Thomas, R.~Moriya, C.~Rettner and
  S.~S.~P.~Parkin, Science {\bf 320}, 209 (2008).

\bibitem{Thomas10} L.~Thomas, R.~Moriya, C.~Rettner, S. and S.~P.~Parkin,
  Science {\bf 330}, 1810 (2010).

\bibitem{Tretiakov10} O.A.~Tretiakov and Ar.~Abanov, Phys.~Rev.~Lett.~{\bf 105}157201 (2010).

\bibitem{Tretiakov12} O.A.~Tretiakov, Y.~Liu and Ar.~Abanov, Phys.~Rev.~Lett.~{\bf 108} 247201 (2012).


\bibitem{SS} V.~Slastikov and C.~Sonnenberg, IMA J.~Appl.~Math.~{\bf 77} no.~2, 220 (2012)

  
\bibitem{HubertSchaefer98} A.~Hubert and R.~Sch\"afer, {\it Magnetic
    Domains: The Analysis of Magnetic Microstructures} (Springer,
  Berlin, 1998).

\bibitem{Sun10} Z.~Z.~Sun and J.~Schliemann, Phys. Rev. Lett. {\bf
    104}, 037206 (2010).

\bibitem{our_PRL} A.~Goussev, J.M.~Robbins, V.~Slastikov,
  Phys. Rev. Lett. {\bf 104}, 147202 (2010).
  
\bibitem{Y_Gou_paper}{  Y.~ Gou, A.~Goussev, J. M.~ Robbins, V.~Slastikov,
Phys. Rev. B {\bf 84}, 104445 (2011)}




    







\bibitem{LandoLifshitz35} L.~D.~Landau and E.~M.~Lifshitz,
  Phys. Zeitsch. Sowietunion {\bf 8}, 153 (1935).

\bibitem{Gilbert55} T.~L.~Gilbert, Phys. Rev. {\bf 100}, 1243 (1955);
  IEEE Trans. Mag. {\bf 40}, 3443 (2004).
  


\bibitem{Kosevich90} A.~M.~Kosevich, B.~A.~Ivanov, and A.~S.~Kovalev,
  Phys. Rep. {\bf 194}, 117 (1990).
  
 \bibitem{paper1} A.~Goussev, R.~Lund, J.M.~Robbins, C.~Sonnenberg, V.~Slastikov, in preparation.


  


 

 











\end{thebibliography}
\end{document}